\documentclass{aastex}

\newcommand{\Mo}{M_{\odot}}

\newcommand{\kms}{km s$^{-1}$}
\newcommand{\Ho}{$H_0$~= 75 \kms\ Mpc$^{-1}$}
\newcommand{\logt}{\log (t/{\rm yr})}

\shorttitle{Mass Function of Star Clusters}
\shortauthors{Zhang \& Fall }
\received{}
\revised{}
\accepted{}

\begin{document}

\title{The Mass Function of Young Star Clusters in the ``Antennae'' Galaxies}

\author{QING ZHANG\altaffilmark{1,2} AND S. MICHAEL FALL\altaffilmark{1} }

\email{qzhang@stsci.edu, fall@stsci.edu}

\altaffiltext{1}{Space Telescope Science Institute, 3700 San Martin Drive, 
                 Baltimore, MD 21218, Email: qzhang@stsci.edu, fall@stsci.edu}
\altaffiltext{2}{Department of Physics and Astronomy, The Johns Hopkins 
       University, 3400 N. Charles Street, Baltimore, MD 21218}

\begin{abstract}
We determine the mass function of young star clusters in the merging 
galaxies the ``Antennae'' (NGC 4038/39) from deep images taken with 
the WFPC2 on the refurbished {\it Hubble Space Telescope (HST)}.
This is accomplished by means of reddening-free parameters and
a comparison with stellar population synthesis tracks to estimate 
the intrinsic luminosity and age and hence the mass of each cluster.
We find that the mass function of the young star clusters (with ages 
$\lesssim 160$ Myr) is well represented by a power law of the form 
$\psi(M) \propto M^{-2}$ over the range $10^4 \lesssim M \lesssim 10^6~ \Mo$. 
This result may have important implications for our understanding of 
the origin of globular clusters during the early phases of galactic 
evolution.
\end{abstract}

\keywords{galaxies: star clusters---galaxies: individual (NGC 4038/39,
the Antennae)---galaxies: interactions}

\section{Introduction}

The luminosity functions of the young star clusters in merging galaxies, 
including the ``Antennae'' (NGC 4038/39),
have roughly power-law form, $\phi(L) \propto L^{\alpha}$ with $\alpha \approx
-2$, and no sign of a peak or turnover down to the limiting magnitudes of 
the observations (Whitmore \& Schweizer 1995; Schweizer et al. 1996;
Miller et al. 1997; Carlson et al. 1998; Whitmore et al. 1999;
Zepf et al. 1999).
These are similar to the luminosity functions of open clusters in the Milky
Way (van den Bergh \& LaFontaine 1984)
and M33 (Christian \& Schommer 1988) and the populous clusters in the LMC
(Elson \& Fall 1985) and starburst
galaxies (Meurer et al. 1995). In contrast, the luminosity functions
of the old globular clusters in the Milky Way, the Andromeda Galaxy, and 
elliptical galaxies have peaks at $L_V \approx 10^5$ $L_{\odot}$ and 
decline toward both higher and lower luminosities. These functions are 
often represented by lognormal distributions of luminosities, corresponding
to Gaussian distributions of magnitudes (e.g., Harris 1991).

This immediately raises the question of whether the mass functions of 
the young star clusters in merging galaxies have power-law or lognormal forms.
The mass functions are of greater relevance to our understanding 
of the physical processes involved in the formation and disruption of 
the clusters. If the star clusters of any population formed at the same
time and with the same stellar initial mass function (IMF), then the 
luminosity 
function of the clusters would always have the same shape as the mass function,
the two being related by a time-dependent shift along the luminosity axis.
However, if the spread in ages within a population is
comparable to the mean age, as seems likely for the young star clusters 
in some recent mergers, one might then worry about the effects of
fading, which could cause the shape of the luminosity function to differ 
substantially from the shape of the mass function (Hogg \& Phinney 1997). 
Based on simulations of this effect, Meurer (1995) concluded that 
the power-law luminosity function in the Antennae derived by 
Whitmore \& Schweizer (1995) from observations with the WFPC1 before the 
refurbishment of {\it HST} may be consistent with an underlying lognormal 
mass function. Furthermore, Fritze-von Alvensleben (1999) estimated the
mass function of young star clusters in the Antennae directly from these
WFPC1 observations and found a lognormal form, similar to that of old 
globular clusters.

The purpose of this Letter is to determine the mass function
of the young star clusters in the Antennae from 
images taken with the WFPC2 after the refurbishment of {\it HST};
these extend about two magnitudes fainter than the earlier images
taken with the WFPC1.
The data reduction and luminosity functions are described in detail 
by Whitmore et al. (1999). 
The novelty of our approach to determine the mass function is that we 
use reddening-free parameters and a comparison with stellar population 
synthesis tracks to estimate the intrinsic luminosity and age and hence
the mass of each cluster.
 
\section{Observations and Selection of Clusters}

The Antennae galaxies were observed with the WFPC2 in 1996 January for 
2000--4500~s with each of the broad-band filters F336W, F439W, F555W, 
and F814W.
Point-like objects were identified and their magnitudes measured with the
IRAF task DAOPHOT; photometry in this system was then transformed to the
Johnson $UBVI$ system.
About 11,000 objects were detected with $17.4 < V < 25.4$,
corresponding to $-14 < M_V < -6$ at the adopted distance of 19.2~Mpc
(for \Ho).
Much of our analysis is based on the reddening-free parameters
\begin{eqnarray}
Q_1=(U-B)-0.72\ (B-V),\\
Q_2=(B-V)-0.80\ (V-I).
\end{eqnarray}
The first of these is the standard reddening-free parameter in the 
$UBV$ system, while the second is a direct extension to the $UBVI$ system. 
Other reddening-free parameters can be defined analogously, but they are 
all linear combinations of $Q_1$ and $Q_2$. Equations (1) and (2) are 
based on the Galactic extinction curve but would be virtually the same
for other familiar extinction curves, such as those of the 
Magellanic Clouds. For objects with $M_V = -9$, the photometric
errors in $U, B, V,$ and $I$ (0.04--0.07) lead to $1\sigma$ uncertainties
of 0.12 in $Q_1$ and 0.11 in $Q_2$. 

We identify cluster candidates and estimate their extinctions and ages 
by comparing them with population synthesis tracks in the $Q_1 Q_2$ diagram, 
shown here as Figure~1.
In the presence of reddening by dust, this approach has significant
advantages over methods based on the more familiar two-color diagrams.
For the population synthesis track, we use the Bruzual \& Charlot (1996, 
hereafter BC96) model with the Salpeter IMF truncated at 0.1 and 
$125~ \Mo$ and with solar metallicity. 
We regard all point-like objects brighter than $M_V = -9$ as cluster 
candidates, irrespective of their $Q$ parameters, because nearly all 
known stars are fainter than this limit (Humphreys 1983).
Furthermore, we regard fainter point-like objects as cluster candidates if 
they lie within $\Delta Q=(\Delta Q_1^2+\Delta Q_2^2)^{1/2}=0.3$ of the 
population synthesis track (except for a few objects with extremely large 
photometric errors).
This procedure eliminates some but not all stars from our sample (see
Figs. 13 and 14 of Whitmore et al. 1999). However, as shown below, 
stellar contamination has little effect on the mass function we 
derive because this is based mainly on the brightest objects. 

For each cluster candidate, we estimate the intrinsic colors, age $t$, and 
mass-to-light ratio $M/L$ from the nearest point on the population synthesis 
track in the $Q_1 Q_2$ diagram.
A comparison of the intrinsic with the observed colors 
gives the extinction $A_V$, and the 
intrinsic luminosity $L_V$ and mass $M$ then follow from 
$V$ and $A_V$. The mean value of $A_V$ varies
from 1.5 for the youngest clusters ($t < 10$ Myr) to 
0.3 for the oldest clusters ($t > 100$ Myr).
Because the BC96 population synthesis track stops evolving in
the $Q_1Q_2$~diagram at $\log (t/{\rm yr}) = 6.4$, no objects can be 
assigned ages smaller than this by our procedure. 
Instead, the youngest clusters are all assigned ages in the 
interval $6.4 < \logt < 6.8$.
Furthermore, because the track bends sharply at $\logt \approx 6.9$ and
has a zigzag at $\logt \approx 7.2$, there is some ambiguity in the assignment
of ages in this region of the $Q_1Q_2$ diagram.
For these reasons, we exclude the intervals $\logt <6.4$ and 
$6.8 < \logt <7.4$ from our determination of the mass function.

We have verified by simulations that the scatter of cluster candidates 
away from the population synthesis track in the $Q_1Q_2$ diagram 
is mostly accounted for by photometric errors.
However, for the youngest objects ($t < 10$~Myr), the scatter is slightly 
larger,
probably because they are affected by nebular emission and residual
extinction. (While our approach is designed to correct for extinction, 
this is never perfect, especially in the dusty regions where the 
youngest clusters are often found.) The errors in the $Q$ parameters
lead to typical uncertainties of a factor of 2 
in the ages and factors that vary from 1.1 to 1.7 in the luminosities. 
Fortunately, these uncertainties tend to cancel out
in estimates of the masses, because younger clusters have lower
mass-to-light ratios but higher extinctions. 
As a result, the uncertainties in the masses of most cluster 
candidates are smaller than 50\%. 
 
Figure~2 shows the luminosity-age relation for the objects  
in our sample along with 
the BC96 population synthesis tracks for $\log (M/\Mo) =$ 6.0, 5.5, 5.0, 
4.5, and 4.0.
Evidently, the masses of the cluster candidates range from
well below $10^4 \Mo$ to just above $10^6~ \Mo$. 
The apparent gaps in the luminosity-age relation at $\logt < 6.4$ and
$7.0 < \logt < 7.2$ are artifacts caused by little or no
evolution of the population synthesis track in the $Q_1Q_2$~diagram 
in these intervals of age (discussed above).
Stellar contamination can be neglected above the horizontal dotted line in 
Figure~2 at $M^c_V=-9$, an extinction-corrected magnitude that excludes all
but the very brightest stars. For reference, the Large Magellanic Cloud, with 
about 4\% of the total blue luminosity of the Antennae, has two stars with 
$M^c_V = -9$, three with $-10 < M^c_V < -9$, and none with $M^c_V < -10$ 
(Humphreys 1983). Thus, we might expect $\sim100$ stars (mostly blue 
supergiants) brighter than $M^c_V = -9$ in our sample of cluster candidates,
corresponding to contamination at the $\sim5\%$ level.

\section{Mass Function of Clusters}

We construct the mass function of cluster candidates in two intervals of age:
$6.4<\logt< 6.8$ and $7.4<\logt< 8.2$,
corresponding to $2.5<t<6.3$~Myr and $25<t<160$~Myr. 
These intervals, which are indicated by the bars at the top of Figure~2, 
were chosen to avoid the problems discussed in the previous section. The first
allows us to estimate the mass function down to relatively low masses
with a relatively large number of objects, 
while the second provides a check on the consistency of our procedure.
In constructing the mass function, we must correct for incompleteness in
our sample. This depends on the brightness of the objects, 
whether they are on the PC or WF chips, and the local background and/or
crowding. For each cluster candidate, we adopt the completeness 
factor determined by Whitmore et al. (1999) using a false-star method.
Our sample as a whole is about 50\% complete at $V=24$, corresponding
to $\log (M/\Mo) \approx 3.9$ for $6.4<\logt<6.8$ and 
$\log (M/\Mo)\approx 4.4$ for $7.4<\logt<8.2$. 
Above these limits, the younger and older subsamples contain 
1140 and 477 objects, respectively.

Figure 3 shows the completeness-corrected mass functions of the cluster 
candidates in the two intervals of age. 
The stellar contamination limits are marked by S's, while the 
completeness limits are marked by C's.
Stellar contamination 
is negligible for the younger subsample, but it could begin to affect
the older subsample for $ \log (M/\Mo) \lesssim 4.7$.
Evidently, the mass functions for the two subsamples are nearly 
indistinguishable above the completeness limits. 
They can be represented by a power law, $\psi(M) \propto M^{\beta}$, with 
$\beta = -1.95 \pm 0.03$ for $6.4<\logt<6.8$ and $\beta = -2.00 \pm 0.08$ 
for $7.4<\logt<8.2$. These are based on weighted least-square fits of
the form $\log\psi = \beta \log M + {\rm const}$. 

We have checked that our results are robust with respect to the adopted 
population synthesis tracks. First, we repeated the entire 
analysis with the BC96 models but with a different IMF 
(Scalo vs. Salpeter) and different metallicities 
($0.4\ Z_{\odot}$ and $2.5\ Z_{\odot}$ vs. $1.0Z_{\odot}$). 
Second, we repeated the analysis with the 
Leitherer et al. (1999) models with the Salpeter 
IMF and solar metallicity. In the first case, the mass function was 
virtually the same; in the second case, it was 
slightly steeper, with $\beta = -2.1$. 
We have also checked that our results are not biased by 
observational errors by randomly re-assigning ages to all cluster candidates
younger than $\logt=6.8$. 
In this case, we obtain $\beta = -1.9$.

Figure 3 shows a comparison between the mass function of the young
star clusters in the Antennae determined here and the mass function of
old globular clusters (indicated by the dash curves).
The latter was derived from the usual Gaussian distribution of magnitudes
and a fixed mass-to-light ratio, $M/L_V = 2$.
Within the observational uncertainties, it appears that the two mass 
functions may be similar for $M \gtrsim 2 \times 10^5~\Mo$. However, for 
$M \lesssim 2 \times 10^5~\Mo$, they are completely different; that for
the young clusters in the Antennae increases rapidly, while that for the old
globular clusters decreases rapidly. 
Whitmore et al. (1999) found that the luminosity function of the star 
clusters in the Antennae could be be described by two power-laws joined 
by a weak bend (with $\alpha_1 = -1.7$ and $\alpha_2=-2.6$),
and there may also be hints of curvature in the mass function we have derived.
However, any such deviations in the latter from a single 
power law have low statistical significance ($<2\sigma$).
Alternatively, the bend in the luminosity function may 
result from the fading of clusters that formed over a period of a few
hundred Myr with a power-law mass function truncated near $10^6~ \Mo$.

Our results differ substantially from those based on earlier observations 
of the Antennae with the WFPC1 on {\it HST} (Meurer 1995; 
Fritze-von Alvensleben 1999).
To understand this difference, we have performed two tests. 
First, we artificially truncated our sample at $M_V = -9.6$, 
the same limit adopted by Fritze-von Alvensleben, and then repeated 
the entire analysis described above. (Note: after corrections for 
extinction, $M_V = -9.6$ corresponds approximately to 
$M^c_V = -9.9$ to $-11.1$). In this case, we obtained a mass function 
similar to the lognormal one found by Fritze-von Alvensleben.
Second, we performed a simulation in which clusters were drawn from
a power-law mass function (with $\beta = -2$) and a uniform age distribution.
The luminosities of the clusters were then computed from the BC96 population
synthesis models. For simulated clusters brighter than $M_V = -9.6$, the
mass function again resembled the one found by Fritze-von Alvensleben.
The reason for this is that, because the clusters fade, 
those with high masses can be observed over a wide range of ages, 
whereas those with low masses can be observed only when they are young.
As a result, low-mass clusters are underrepresented in the observed
mass function, which therefore declines toward both high and low masses.

\section{Discussion}

We have found that the mass function of young star clusters 
in the Antennae is well represented by a power law,
$\psi(M) \propto M^{-2}$, over the range $10^4 \lesssim M \lesssim 10^6~ \Mo$.
This is similar to the power-law mass function of diffuse and
molecular clouds in the Milky Way (Dickey \& Garwood 1989;
Solomon \& Rivolo 1989). However, it differs radically from 
the lognormal mass function of old globular clusters, which 
peaks at a ${\rm few}\times10^5~\Mo$ and declines rapidly
toward both higher and lower masses. 
It is widely believed that galaxies formed hierarchically by the merging 
of smaller galaxies and/or subgalactic fragments. Thus, our results have 
potentially important implications concerning the origin of globular 
clusters during the early phases of galactic evolution.
In this connection, it is worth emphasizing that a power-law mass 
function is ``scale free,'' whereas a
lognormal mass function has a ``preferred scale.''

One explanation for the different mass functions is that the conditions
in ancient galaxies and protogalaxies were such as to imprint a 
characteristic mass of a ${\rm few}\times10^5~\Mo$ but that these conditions
no longer prevail in modern galaxies. For example, the minimum mass 
of newly formed star clusters, set by the Jeans mass of interstellar clouds, 
will be high when the gas cannot cool efficiently and low when it can, which
in turn will depend on the abundances of heavy elements and molecules, the
strength of any heat sources, and so forth. These effects favor the formation 
of clusters in a narrower range of masses in the past than at present 
(Fall \& Rees 1985; Kang et al. 1990). 
Another explanation for the different mass functions is that populations 
of stars clusters were born scale free, but later acquired a preferred
scale by the selective disruption of low-mass clusters (Fall \& Rees 1977;
Gnedin \& Ostriker 1997, and references therein). 
In this case, a power-law mass function might evolve into a lognormal mass 
function (Okazaki \& Tosa 1995; Elmegreen \& Efremov 1997; Baumgardt 1998;
Vesperini 1998). Whether this occurs, however, depends on other aspects of
the initial conditions, especially how the mass-radius plane is populated.
We plan to address this issue in a future paper. 

\acknowledgements
We thank Claus Leitherer and Brad Whitmore for valuable discussions. Support
for this work was provided by NASA through grant number GO-07468 from STScI.

\newpage
\figcaption{
$Q_1 Q_2$ diagram. The line represents the BC96 population synthesis
track with the indicated values of $\log (t/{\rm yr})$ and tick marks 
every $\Delta \log t = 0.2$. In this model, the $Q$ parameters remain 
constant for $\logt < 6.4$. The dots are all the cluster candidates
in our sample with estimated masses above $10^5~ \Mo$. 
}

\figcaption{
Extinction-corrected luminosities of cluster candidates as a 
function of their age. The lines represent the BC96 population 
synthesis tracks with $\log (M/\Mo) =$ 6.0, 5.5, 5.0, 4.5, and 4.0. 
The bars at the top indicate the intervals of age adopted to determine 
the mass function. The horizontal dotted line at $M^c_V = -9$ 
(extinction-corrected)
indicates the upper limit of stellar contamination.
}

\figcaption{
Completeness-corrected mass function $\Psi(\log M)= M \psi(M)$ of 
cluster candidates in two intervals of age:
$6.4<\logt<6.8$ {\it (top)} and $7.4<\logt<8.2$ {\it (bottom)}.
The vertical and horizontal bars indicate the $1\sigma$ Poisson 
uncertainties and the bin sizes, respectively.
The arrows with S indicate the stellar contamination limit ($M^c_V\approx-9$),
and the arrows with C indicate the $50\%$ completeness limit ($V\approx 24$).
The straight lines are power laws, $\psi(M)\propto M^{\beta}$, with
$\beta=-1.95\pm 0.03$ and $\beta=-2.00\pm 0.08$, respectively,
for the younger and older clusters.
The dashed line represents a lognormal mass function, derived from
a Gaussian distribution of magnitudes with a mean $\left<M_V\right>= -7.3$
and dispersion $\sigma (M_V)= 1.2$ (Harris 1991)
and a fixed mass-to-light ratio $M/L_V=2$.
}

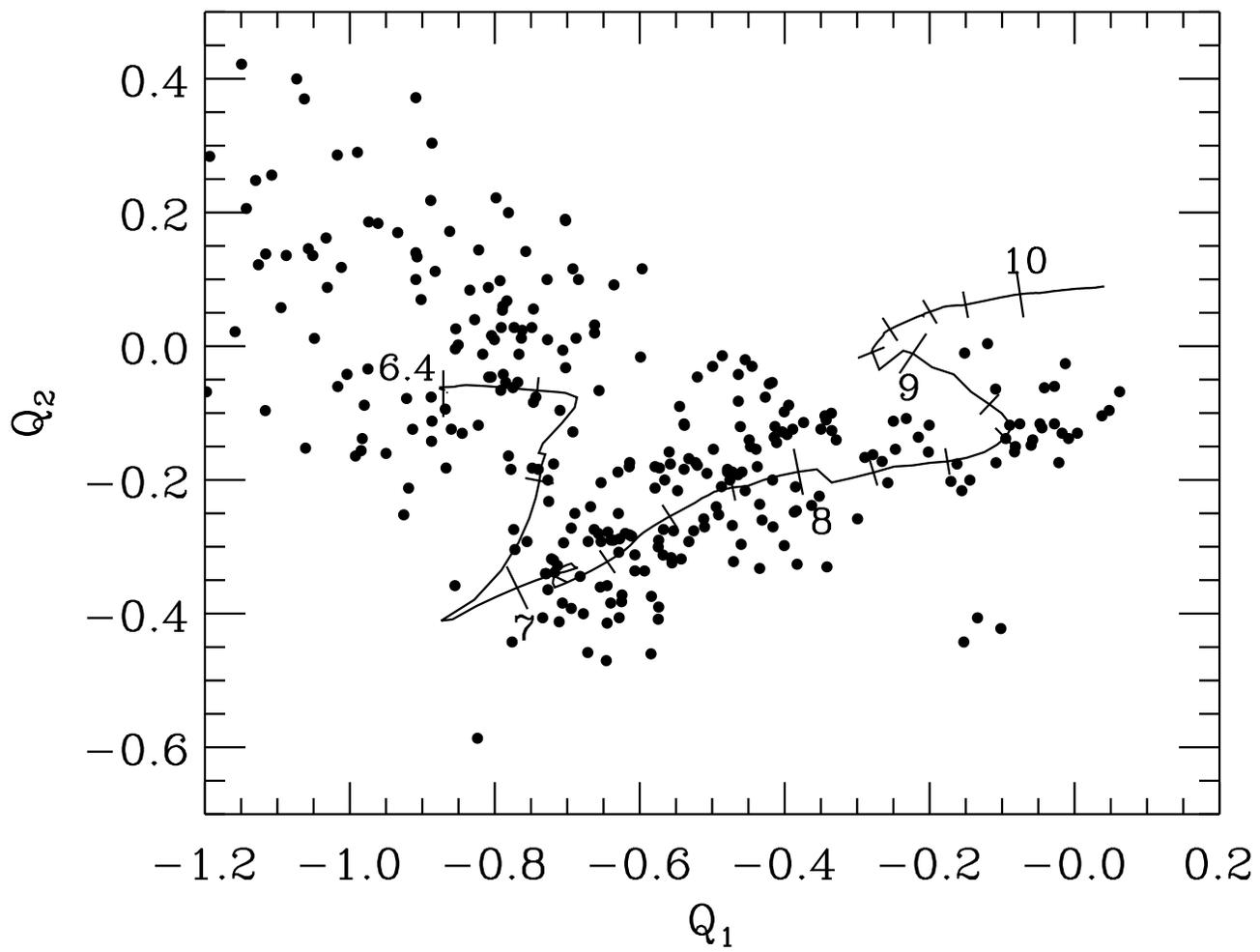
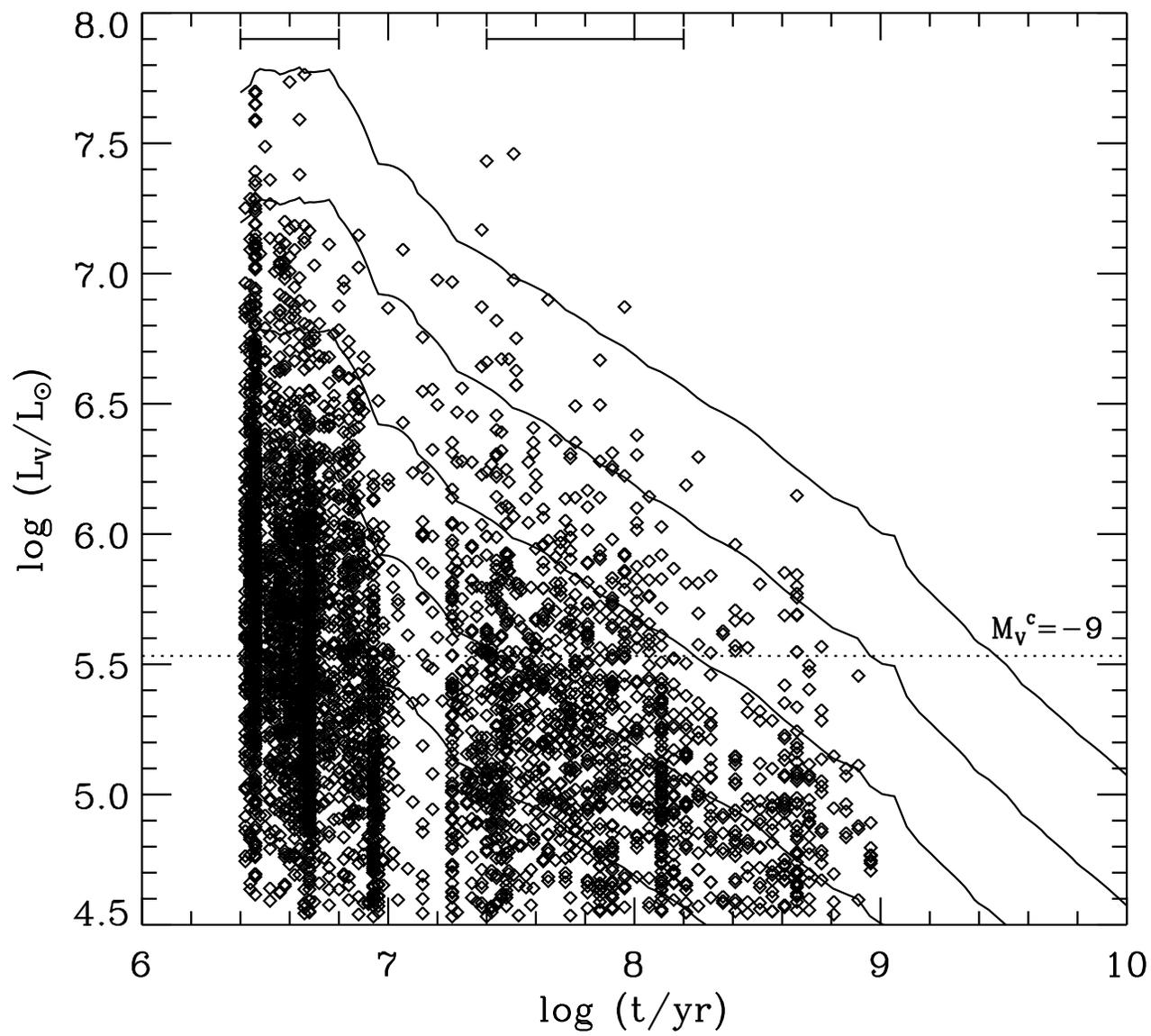
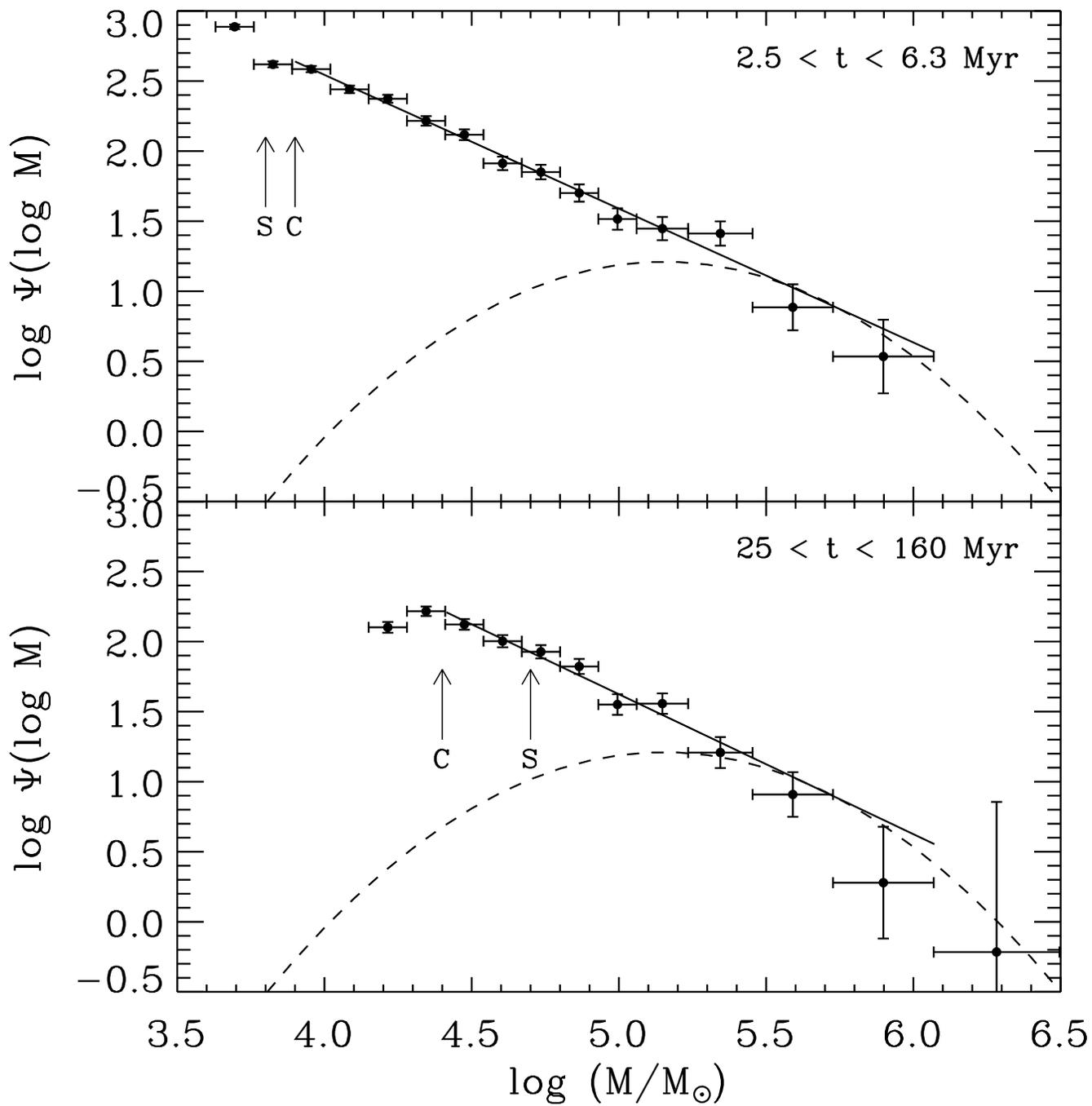
\end{document}